\documentclass[
]{ceurart}

\usepackage{listings}
\usepackage[normalem]{ulem}
\useunder{\uline}{\ul}{}

\usepackage{graphicx} 
\usepackage{tabularx}
\usepackage{subcaption} 
\usepackage{amsthm}
\usepackage{xcolor}
\usepackage{multirow}
\usepackage{booktabs}
\usepackage{longtable}
\usepackage{placeins}
\usepackage{wrapfig}

\newtheoremstyle{mystyle} 
  {10pt}                   
  {10pt}                   
  {\itshape}              
  {}                       
  {\color{blue}\bfseries} 
  {.}                      
  { }                      
  {}                       
\theoremstyle{mystyle}

\usepackage{microtype}

\usepackage{inconsolata}
\usepackage{amsmath}
\usepackage{tcolorbox}

\tcbset{
  nobox/.style={
    boxrule=0pt,
  }
}

\tcbset{
  compactbox/.style={
    boxrule=0pt,
    colframe=white,
    colback=white,
    coltitle=blue!40!black,
    coltext=black,
    sharp corners,
    boxsep=2pt,
    left=2pt,
    right=2pt,
    top=2pt,
    bottom=2pt,
  }
}

\definecolor{MyRed}{rgb}{1,0,0}

\definecolor{MyBlue}{rgb}{0,0,1}

\definecolor{MyDarkOrange}{rgb}{0.8, 0.4, 0}

\begin{document}

\copyrightyear{2025}
\copyrightclause{Copyright for this paper by its authors. Use permitted under Creative Commons License Attribution 4.0 International (CC BY 4.0).}

\conference{ReNeuIR 2025 (at SIGIR 2025) -- 4th Workshop on Reaching Efficiency in Neural Information Retrieval, July 17, 2025, Padua, Italy}

\title{Drowning in Documents: Consequences of Scaling Reranker Inference}


\author[1,2]{Mathew Jacob}
\fnmark[1]
\fntext[1]{Work done during Mathew's internship at Databricks.}

\address[1]{Databricks, San Francisco, USA}
\address[2]{University of Illinois at Urbana-Champaign, Urbana, USA}

\author[1]{Erik Lindgren}

\author[1]{Matei Zaharia}

\author[1]{Michael Carbin}

\author[1]{Omar Khattab}

\author[1]{Andrew Drozdov}[
email=andrew.drozdov@databricks.com,
]
\cormark[1]

\cortext[1]{Corresponding author.}

\begin{abstract}
Rerankers, typically cross-encoders, are computationally intensive but are frequently used because they are widely assumed to outperform cheaper initial IR systems. We challenge this assumption by measuring reranker performance for full retrieval, not just re-scoring first-stage retrieval. To provide a more robust evaluation, we prioritize strong first-stage retrieval using modern dense embeddings and test rerankers on a variety of carefully chosen, challenging tasks, including internally curated datasets to avoid contamination, and out-of-domain ones. Our empirical results reveal a surprising trend: the best existing rerankers provide initial improvements when scoring progressively more documents, but their effectiveness gradually declines and can even degrade quality beyond a certain limit. We hope that our findings will spur future research to improve reranking.
\end{abstract}


\maketitle

\section{Introduction}

Contemporary information retrieval (IR) models are often either \textbf{retrievers} \cite[inter alia]{reimers-gurevych-2019-sentence,karpukhin-etal-2020-dense,izacard2022unsupervised} which pre-compute document representations for efficient retrieval or \textbf{rerankers} \cite[inter alia]{Nogueira2019PassageRW,gao2021multistage,glass-etal-2022-re2g} that jointly encode query--document pairs, often with cross-encoder architectures. These two paradigms pose a popular tradeoff between quality and cost: retrievers are orders of magnitude cheaper as they pre-index document representations, and while rerankers are more expensive, they are widely understood in the literature to boost quality and generalization due to their expressive modeling capabilities~\cite{Rosa2022InDO}.

Modern IR systems often manage this tradeoff using \textit{multi-stage reranking pipelines} \cite{matveeva2006nested,lidan2011cascade,nogueira2019multi}, where a fast retriever identifies the initial top-$K$ candidate documents and a reranker then re-scores only those $K$.
If the models used are well-chosen---e.g. best-in-class retrievers and rerankers---it is generally assumed both (1) \textit{that introducing rerankers will consistently improve overall quality} and (2) \textit{that increasing the number of reranked documents $K$ will progressively lead to even higher gains}~\cite[inter alia]{Humeau2020Poly-encoders,luan-etal-2021-sparse}.
As a corollary, using rerankers to score the entire document set should also be an effective---albeit potentially unrealistic in terms of cost---approach for achieving high recall.





\begin{figure*}[t!]
    \centering
    \begin{subfigure}[t]{0.48\linewidth}
        \centering
        \includegraphics[width=\linewidth]{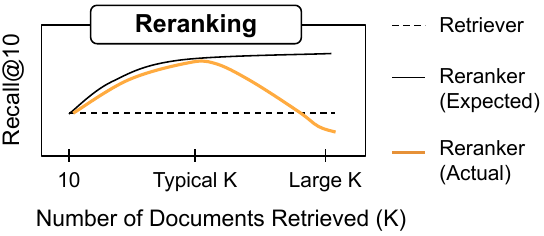}
    \end{subfigure}
    \hfill
    \begin{subfigure}[t]{0.48\linewidth}
        \centering
        \includegraphics[width=\linewidth]{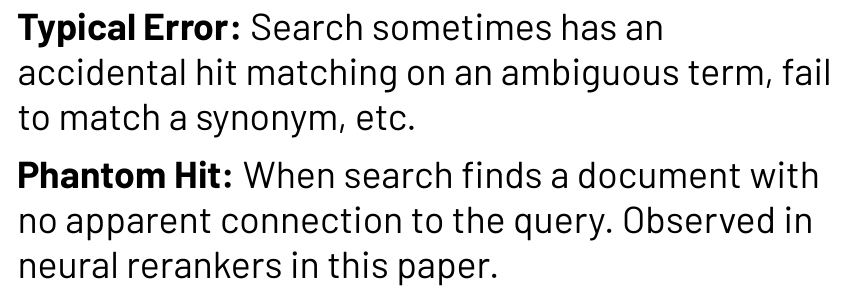}
    \end{subfigure}
    \caption{Left: While rerankers are thought to outperform retrievers, we find instances where scaling the number of documents $K$ used in reranking often leads to a substantial decrease in recall.
    Right: Our definition of \textit{phantom hit}, which is an unexpected error we observed in various neural rerankers.}
    \label{fig:simple_fig_rerank}
    \vspace{-4mm}
\end{figure*}




We test these seemingly innocuous assumptions using best-in-class rerankers across several underexplored public and enterprise IR benchmarks.
Notably, our choice of datasets and models departs from conventional reranker research, which typically focuses on BM25 for first-stage retrieval and MSMarco for evaluation.
We find that these assumptions are frequently false. Specifically, while rerankers initially help with small values for $K$, reranking with large $K$ decreases recall precipitously (Figure~\ref{fig:simple_fig_rerank}), often dropping beneath the quality of standalone retrievers. 
As a consequence, modern rerankers frequently perform \textit{worse} than retrievers when both rank the full dataset. Upon studying this, we find that rerankers often score completely irrelevant documents, lacking semantic or lexical similarity to the query, very highly. These same documents, which we call \textit{phantom hits}, are often given very low scores by the retrievers, revealing a counterintuitive way in which retriever models can outperform reranker models in practice.

Our main findings are thus that (1) we empirically demonstrate that \textbf{scaling K leads to a substantial decrease in recall} in multiple settings, (2) we qualitatively reveal and investigate \textbf{a pathology in reranking that we describe as Phantom Hits}, where the reranker assigns a high score to documents that have no lexical or semantic overlap with the query, and (3) we demonstrate a proof-of-existence that \textbf{more robust reranking options exist} and can be implemented with large language models (LLMs).
Overall, we establish that the current understanding of rerankers does not match their behavior in practice and call for more research to improve their robustness to noise.

\section{Background and Related Work}

\paragraph{Retrievers}
A first-stage retriever takes a query and searches the entire corpus to find relevant documents. Retrievers usually embed the documents offline and compute a cheap similarity score between the embedded query and documents to find the most relevant documents. For sparse vector methods, BM25 \cite{original_bm25} has consistently proven to be a strong baseline. BM25 is a bag-of-words method based on lexical matching that uses an inverted index for fast search.

For dense retrieval, Transformers are used to encode the query and document separately
\cite{reimers-gurevych-2019-sentence, karpukhin-etal-2020-dense, izacard2022unsupervised}. 
This degree of independence enables dense embeddings to be computed for each document offline. At search time, only the query needs to be encoded, and then vector search algorithms are used to quickly find the most relevant documents \cite{johnson2019billion, malkov2018efficient, avq_2020}.

\paragraph{Rerankers} A \textit{cross-encoder} \cite{Nogueira2019PassageRW, gao2021multistage, glass-etal-2022-re2g} is a model that, given a query and document, outputs a relevance score for the pair. The cross-encoder attends to the query-document pairs \textit{jointly}. The highly-expressive modeling of cross-encoders is widely understood in the literature \cite{Humeau2020Poly-encoders,Rosa2022InDO} to lead to much better accuracy and generalization. 

\paragraph{Scaling Compute} A growing body of work investigates the effects of scaling compute on different components of IR and RAG systems. \citet{scalinglaws2024retrieval} investigated the scaling laws of dense retrieval models, measuring how factors like model and data size affect performance of neural retrievers. \citet{2024LongContextRAG} and \citet{longcontextscalinginference} explore improving long-context RAG performance by scaling inference compute. 
There is also a broader body of work on scaling compute for LLM training \cite{Kaplan2020ScalingLF} and inference \cite{largelanguagemonkeys, gpto1, scalinginferencereasoning}.

\section{Experimental Setup}

We test (1) how first-stage retrievers and rerankers interact across different model pairings and reranking depth $K$ and (2) to compare retrievers against rerankers for full dataset retrieval.

\subsection{Retrievers}  

We consider various retrievers with different cost and quality tradeoffs. For simplicity and reproducibility we use exact scoring for all retrievers. Refer to Table \ref{tab:model_links} in the Appendix for more information on retrievers.

    \paragraph{Lexical Search} BM25 is a lexical search approach that sparsely represents each document according to their token counts. We use the implementation from Pyserini \citep{Lin_etal_SIGIR2021_Pyserini} with the default parameters. BM25 is fast at scale when using inverted indices, although can not represent semantic similarity between text.

    \paragraph{Dense Embeddings} We use two proprietary embedding models, namely, \texttt{voyage-2}, a 1024-dimensional dense embedding model from Voyage AI, and \texttt{text-embedding-3-large}, a 3072-dimensional dense embedding model from OpenAI. The latter is the highest quality but most costly retrieval model that we consider.

\subsection{Rerankers} 

We study several state-of-the-art open and closed rerankers. Refer to Table \ref{tab:model_links} in the Appendix for more information on rerankers.

    \paragraph{Open Models} We include two high quality open source cross-encoder models in our experiments: \texttt{jina-reranker-v2-base-multilingual} and \texttt{bge-reranker-v2-m3}. Both rerankers are published on huggingface. We access the Jina model through its API and run BGE locally. 

    \paragraph{Closed Models} We include rerankers from Cohere and from Voyage AI: \texttt{rerank-english-v3.0} and \texttt{voyage-rerank-1}. While these models have been described as cross-encoders in various blog posts, we cannot verify the precise architecture of closed models. We access the rerankers through their respective APIs.

\subsection{Datasets}

For our evaluation, we use a combination of eight academic and enterprise datasets to ensure adequate coverage of realistic retrieval workloads.

\paragraph{Academic Datasets} Our experiments include evaluations across five diverse academic datasets.
We include the biology and pony splits from BRIGHT \cite{BRIGHT} that require \textit{reasoning} beyond lexical and semantic matching, the relic and doris-mae splits from BIRCO \cite{Wang2024BIRCOAB} that have been pathologically filtered down to the most challenging queries, and also scifact from BEIR \cite{thakur2021beir}, one of the most well established leaderboards for dense embeddings.

\paragraph{Enterprise Datasets} 
To capture workloads that are representative of industry use cases, we benchmark rerankers across three internally curated datasets.
FinanceBench is a RAG dataset derived from \citet{financebench}, and includes company-specific questions answerable from SEC filings.
ManufacturingQA is derived from an internal knowledge base over technical documentation about manufacturing. This dataset is representative of real domain-specific customer queries and may contain alphanumeric product codes.
DocumentationQA is an internal dataset that is comprised of users' code-related questions with manually labeled answers grounded in documentation from an open-source software framework.

\paragraph{Additional Details}


We make two simplifications when preprocessing  datasets. First, we truncate queries and documents to 512 tokens so that embeddings and rerankers are on a similar playing field, since many research papers have pointed out the challenges for long context retrieval with embeddings \cite{zhu-etal-2024-longembed}. Second, we downsample negative documents to a maximum of $N=10,000$ documents---this reduces costs of experiments, and increasing N would only show a more profound version of existing trends. 
Additionally, for the enterprise datasets we were unable to evaluate against \texttt{jina-reranker-v2-base-multilingual} due to legal constraints.
We provide a brief guide to ease reproducibility in Appendix~\ref{app:reproducibility}.

\subsection{Why these datasets and models?}

If we focused exclusively on weak retrieval, then we would have a false impression about reranker performance. For example, even though BM25 is well known to have impressive generalization capabilities, recent dense embeddings can greatly outperform BM25 across multiple benchmarks.

If we focused exclusively on the most popular benchmarks for reranker evaluation, such as MSMARCO and BEIR, then similarly we would have a false impression about reranker performance. Rerankers have been directly tuned for these evaluations either directly through including relevant samples in their training, or indirectly through validation. Our diverse selection of datasets not only covers challenging settings that rerankers have not been optimized for, it also is more representative of the real world setting where off-the-shelf models are used on enterprise data. In future work, we hope to investigate even larger corpora that contain tens or hundreds millions of passages such as TREC-RAG~\cite{Pradeep2024RagnarkAR} (despite its reliance on MSMARCO) and TREC-Biogen~\cite{Gupta2024OverviewOT}.





\begin{figure*}[b!]
    \centering
    \includegraphics[width=1\linewidth]{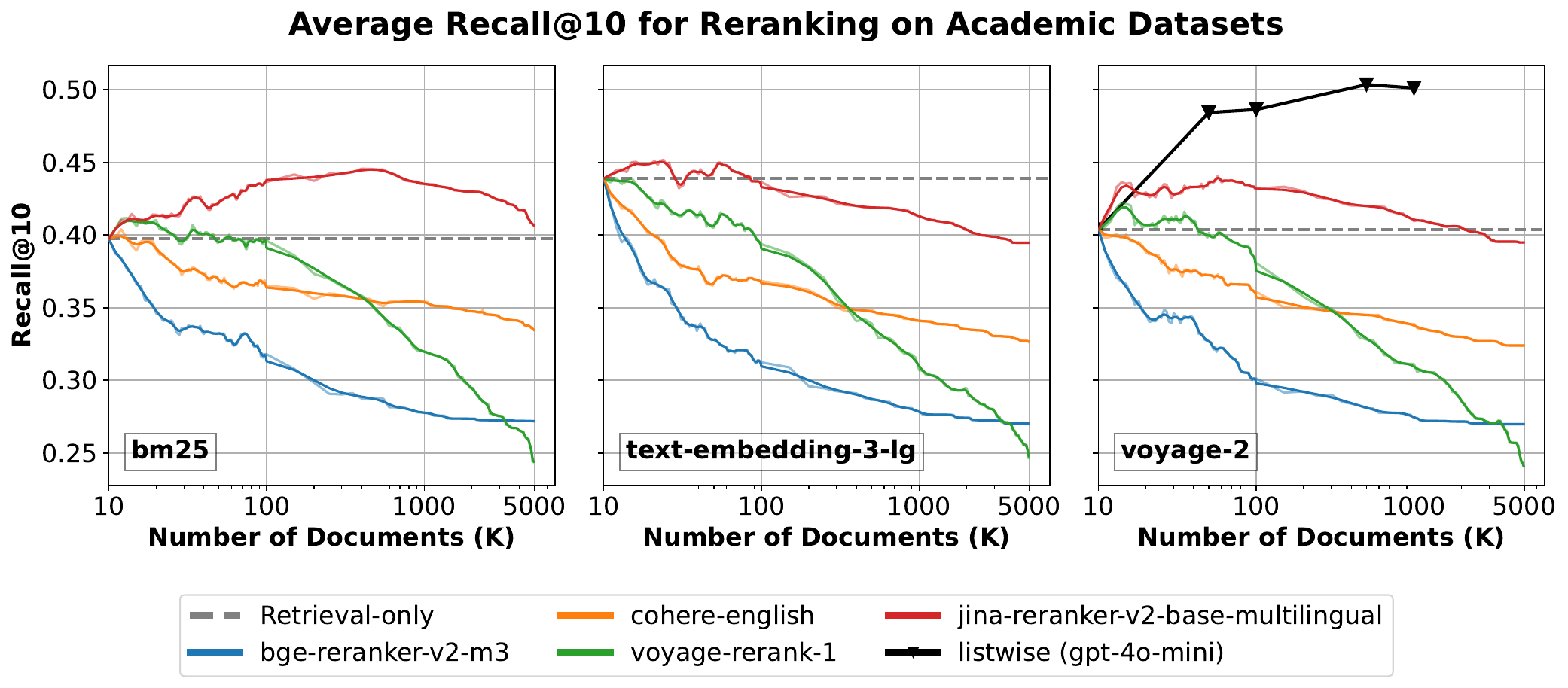}
    \caption{Recall@10 when reranking, averaged across academic datasets. The dashed line shows the first stage recall, and the solid line is the rerankers' recall. The rerankers' recall often degrades as the reranked $K$ increases.}
    \label{fig:recall_10_plots}
\end{figure*}

\begin{figure*}[t!]
    \includegraphics[width=1\linewidth]{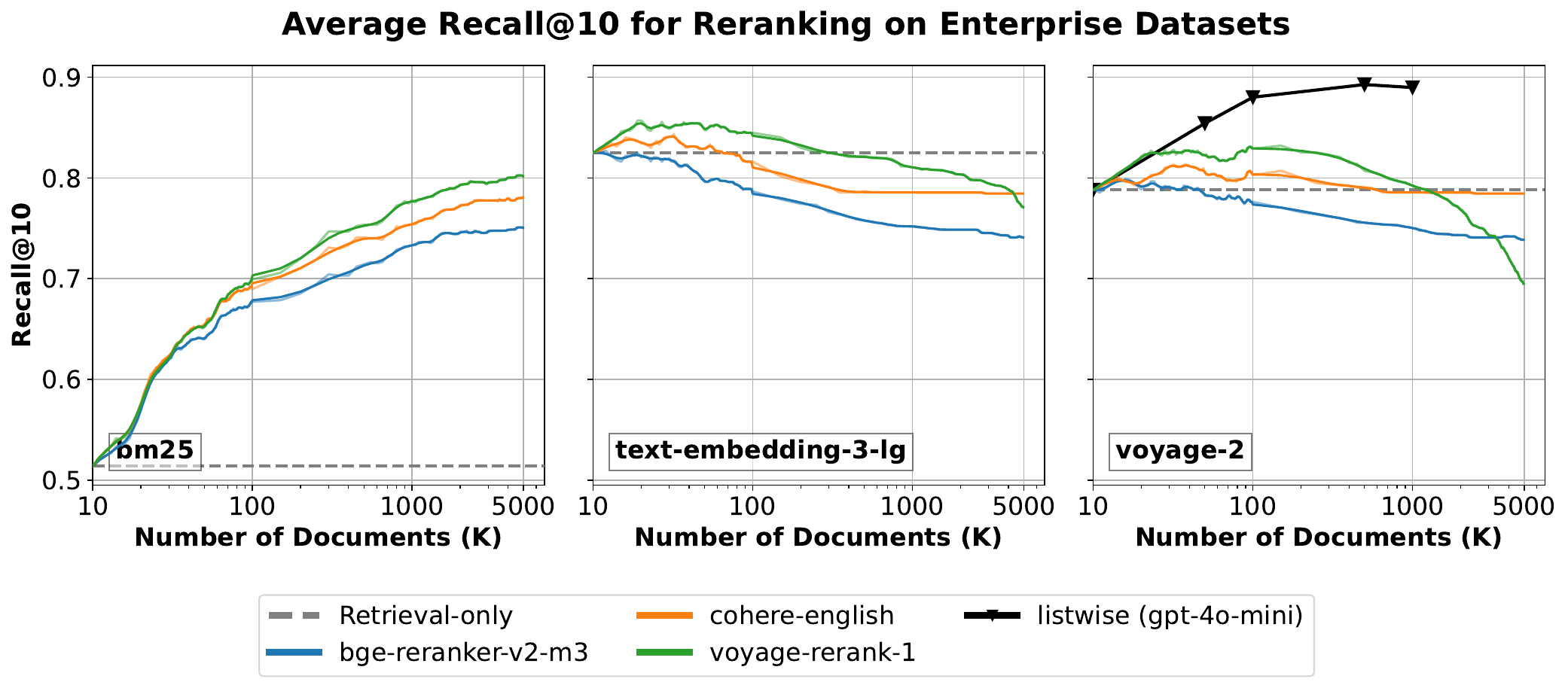}
    \caption{Recall@10 when reranking, averaged across enterprise datasets. The dashed line shows the first stage recall, and the solid line is the rerankers' recall. The rerankers' recall often degrades as the reranked $K$ increases.}
    \label{fig:recall_10_plots_enterprise}
\end{figure*}

\section{Results \& Analysis: Scaling Up Reranking \label{sec:rerank}}

How well do modern rerankers perform when given different amounts of documents to rerank? We measure quality when reranking the top-$k$ documents from different retrievers. We vary $k$ to sizes much larger than previous evaluations (larger than 5000)\footnote{For cross-encoder evaluations, reranking the top-100 \cite{gao2021multistage} or top-1000 \cite{Zhuang2022RankT5FT,xenc_query_expansion_2024} documents is typical.} to better understand how rerankers behave in extreme settings. Figures \ref{fig:recall_10_plots} and \ref{fig:recall_10_plots_enterprise} report Recall@10 for the rerankers, averaging across enterprise and academic datasets respectively. 
In the Appendix~\ref{app:detailed_results}, we include finer-grained results on the individual datasets. 

\subsection{Does scaling the number of documents for reranking help? \label{sec:scaling_helps}}

In the majority of cases, we see that reranking a small set of documents is an effective way to improve recall on a dataset (Figures~\ref{fig:recall_10_plots} and \ref{fig:recall_10_plots_enterprise}). This claim is reinforced by the statistics from individual experiments,\footnote{An experiment is uniquely identified by a (dataset, retriever, reranker) tuple.} which show that rerankers improved Recall@10 over retrieval for at least one value of $K$ in 85.0\% and 88.9\% of the time across academic and enterprise datasets, respectively. The full results for individual experiments are included in Appendix~\ref{app:detailed_results}.

A closer look at the results reveals that even though rerankers are effective when reranking a few documents, that \textbf{rerankers are much less consistently helpful when scaling the number of documents for reranking}. To make this more clear, in Table~\ref{tab:analysis_scaling} we include a fine-grained analysis of reranker average Recall@10 at different values of K by filtering for different cases of interest. Our initial filter (\textbf{Helps}) is to look for experiments where there exists at least one K such that reranker Recall@10 is greater than the retriever's. We investigate four additional filters (\textbf{Helps + X}):

\begin{itemize}
    \item \textbf{Never Hurts:} There is no K such that reranker Recall@10 is less than the retriever's.
    \item \textbf{Distracted:} The reranker improves over the retriever at K, but there is K' > K where the reranker's Recall@10 for K' is less than its Recall@10 for K.
    \item \textbf{Scaling is Best:} The reranker's maximum Recall@10 is at the maximum K.
    \item \textbf{Scaling Hurts:} The reranker's Recall@10 at the maximum K is worse than the retriever's.
\end{itemize}

\begin{table}[h!]
    \centering
\begin{tabular}{lrrrrrrrrrr}
\toprule
{} & \multicolumn{6}{c}{Academic} & \multicolumn{4}{c}{Enterprise} \\
\cmidrule(lr){2-7} \cmidrule(lr){8-11}
{} &   SCI &     DM &    REL &    BIO &   PONY &  All &    FIN &    MQA &    DQA &  All \\
\midrule
\# of Experiments                       &          12 &          12 &           12 &         12 &          12 &          60 &           9 &           9 &          9 &          27 \\
\midrule
Helps                   &  83.3 &  91.7 &  100.0 &  66.7 &  83.3 &  85.0 &  100.0 &  100.0 &  66.7 &  88.9 \\
Helps \& Never Hurts     &  25.0 &   8.3 &   75.0 &   0.0 &   8.3 &  23.3 &   11.1 &   22.2 &  33.3 &  22.2 \\
Helps \& Distracted      &  50.0 &  83.3 &  100.0 &  16.7 &  50.0 &  60.0 &   88.9 &   66.7 &  44.4 &  66.7 \\
Helps \& Scaling is Best &   8.3 &   8.3 &    0.0 &   0.0 &   0.0 &   3.3 &   11.1 &   22.2 &  11.1 &  14.8 \\
Helps \& Scaling Hurts   &  41.7 &  66.7 &   16.7 &  66.7 &  75.0 &  53.3 &   44.4 &   55.6 &  33.3 &  44.4 \\
\bottomrule
\end{tabular}
    \caption{Experiment-level statistics for Recall@10 when reranking. We filter for series using the base condition (Helps) and the four additional filters described in \S\ref{sec:scaling_helps}. Each cell shows the percentage (\%) of experiments for which the conditions are met.}
    \label{tab:analysis_scaling}
\end{table}

\textbf{One of our key observations is that continuing to include more documents for the reranker is an ineffective way to scale computation.} In very few cases did using the largest value of $K$ lead to to the best average Recall@10 (see ``Helps \& Scaling is Best'' in Table~\ref{tab:analysis_scaling}). Furthermore, 53.3\% and 44.4\% of the time across academic and enterprise datasets, respectively, we saw that even when reranking a small $K$ was effective, that scaling $K$ lead to Recall@10 that was worse than retrieval alone (see ``Helps \& Scaling Hurts'' in Table~\ref{tab:analysis_scaling}).


On the enterprise data, reranking the documents from BM25 always lead to an improvement, plus, reranking consistently seems to improve as the number of documents increases. \textbf{BM25 plus reranking might seem promising, but the reality is that this can be outperformed by retrieval-only with strong dense embeddings.}
For example, we see that text-embedding-3-large is roughly 1.5x as effective as BM25 on the enterprise data, measured by average Recall@10. Retrieving a few documents with a dense embedding would be better than reranking many documents with BM25.\footnote{Our focus on strong retrievers and challenging datasets is in contrast to previous work that relies heavily on BM25 and MS MARCO~\cite{hof2019lowfreq}.} The stark contrast between BM25 and the dense embeddings is another example of the challenges for retrieval systems to generalize, and the necessity of many varied evaluation suites.

\subsubsection{How to fairly compare retrieval and reranking}\label{sec:full_rerank}

Given that rerankers often show improved retrieval quality when reranking a few documents, one may assume that rerankers are more accurate than retrievers. Additionally, it's rarely clear how effective a reranker really is, since a reranker is usually evaluated jointly with first stage retrieval. With this in mind, we develop a fair and simple protocol for comparing retrievers and rerankers: Rather than apply the reranker only on the top documents provided by the retriever, apply the reranker on \textit{all} of the documents that are being retrieved from.

\begin{figure}[h]
    \centering
    \includegraphics[width=1\linewidth]{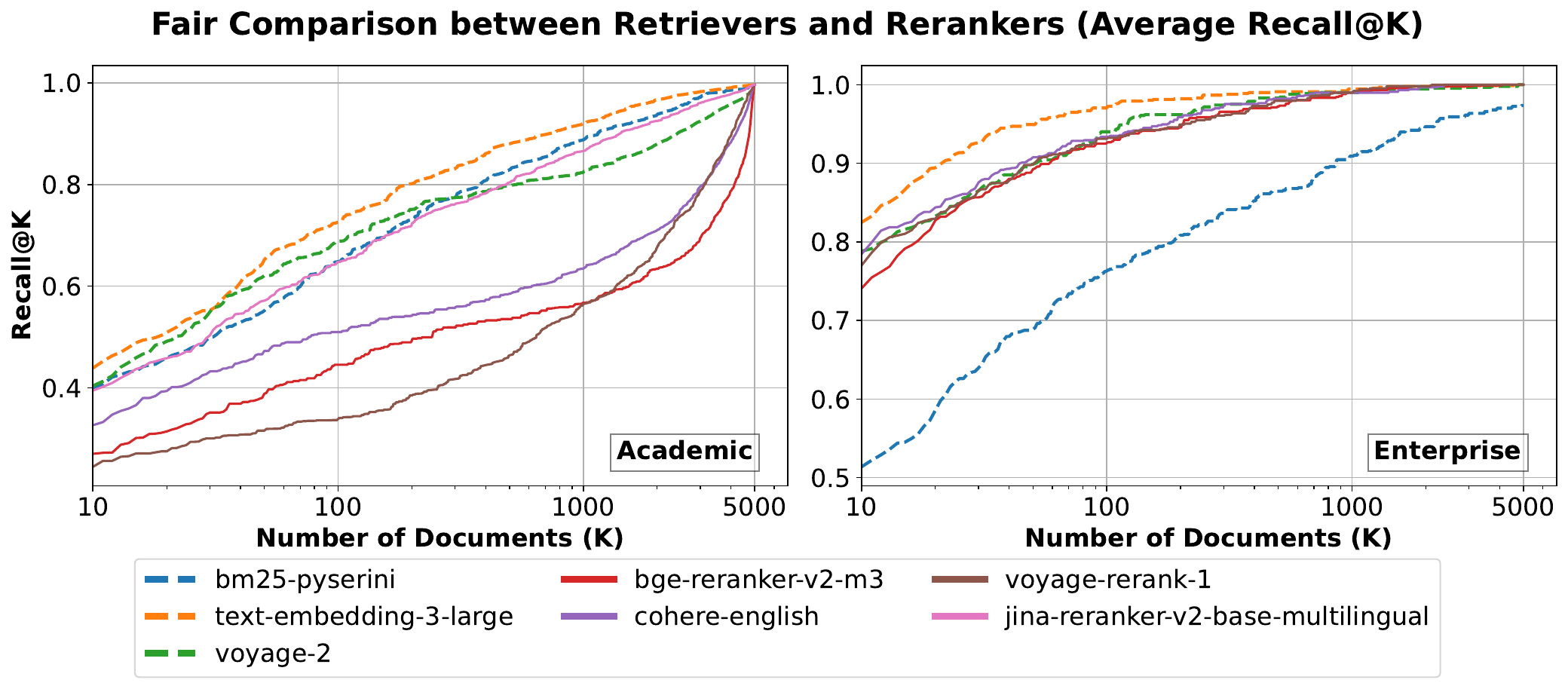}
    \caption{Recall@K for full retrieval, averaged across academic and enterprise datasets. The maximum recall can be less than 1 since we only display recall up to 5000 documents, and some datasets contain additional documents. Note the difference in y-axis.}
    \label{fig:recall_k}
\end{figure}

We use this protocol with our retrievers and rerankers of interest and present the results in Figure~\ref{fig:recall_k}, where we plot Recall@K across all values of K. In line with our previous findings, we see that retrievers are better than rerankers in many settings.

\subsection{Reranker Errors and \textit{Phantom Hits}\label{sec:error_analysis}}


In Table~\ref{tab:analysis_scaling}, we see that while reranking with a small number of documents can be beneficial, reranking with a large K often leads to worse performance: Recall@10 is lower than that of retrieval alone in 53.3\% and 44.4\% of experiments on academic and enterprise datasets, respectively. In a larger percentage of cases we find helpful rerankers will get distracted and Recall@10 will dip at higher K, though not always falling below the retriever.
It is plausible that this decrease in recall could be due to missing labels, and rerankers finding previously unknown but relevant documents. However, when manually looking through the reranker's predictions, we see that this is not the case. Instead, we attribute many reranker errors to what we call ``phantom hits''. These are cases where the reranker prefers an irrelevant negative document to the positive ground truth when there is no apparent connection between the query and irrelevant document, lexical, semantic, or otherwise. In this error analysis, we seek a few representative examples of these ``phantom hits''.

To mine potentially interesting examples, we filtered for queries where the Recall@10 decreased when reranking K=5000 instead of K=100 documents. We show examples of these reranker failures in Figures~\ref{fig:error_analysis_1} and \ref{fig:error_analysis_2}. In all the selected cases, the retriever (text-embedding-3-large) assigns a higher rank to the shown positive document, sometimes by a large margin. However, we see multiple cases where the reranker prefers irrelevant \textit{phantom hits} over positive documents even when they have little or no text overlap with the query. We found this behavior is not isolated to a single reranker, and that seemingly random documents about ``dishwashing'' and ``exercise'' were preferred when there were clearly more topically similar options. More detailed rankings of each query are shown in Appendix~\ref{app:error_analysis}.

\begin{figure}[bh!]
\begin{subfigure}[b]{\linewidth}
\begin{tcolorbox}[colframe=black, colback=white, boxrule=0.5pt, width=\linewidth, boxsep=5pt, left=5pt, right=5pt,
title={\small\centering text-embedding-3-large + voyage-rerank-1}]
    {\small 
\textbf{Query} \\ Many \textcolor{blue}{proteins} in human cells can be post-translationally modified at \textcolor[rgb]{0.13, 0.55, 0.13}{lysine} residues via \textcolor{orange}{acetylation}.
}

\vspace{2mm}\hrule\vspace{3mm}

{\small
\textbf{Positive Document} \textbf{[Retriever: 1, Reranker: 39]} \\
\textcolor{blue}{Protein} \textcolor[rgb]{0.13, 0.55, 0.13}{Lysine} \textcolor{orange}{Acetylated}/Deacetylated Enzymes and the Metabolism-Related Diseases \textcolor[rgb]{0.13, 0.55, 0.13}{Lysine} \textcolor{orange}{acetylation} is a reversible posttranslational ...
}

\vspace{2mm}\hrule\vspace{3mm}

{\small
\textbf{Negative Document} \textbf{[Retriever: 3337, Reranker: 5]} \\
On the origins of ultra-fine anaphase bridges. Comment on: Chan KL, Palmai-Pallag T, Ying S, Hickson ID. Replication stress induces sister-chromatid bridging at fragile site loci in mitosis. Nat Cell Biol \textquotesingle09; 11:753-60.
}
\end{tcolorbox}
\end{subfigure}

\vspace{4mm}

\begin{subfigure}[b]{\linewidth}
\begin{tcolorbox}[colframe=black, colback=white, boxrule=0.5pt, width=\linewidth, boxsep=5pt, left=5pt, right=5pt,
title={\small\centering text-embedding-3-large + bge-reranker-v2-m3}]
\input{examples/bge-binding}
\end{tcolorbox}
\end{subfigure}

    \caption{Two unexpected reranker error from Scifact. The ranks assigned by the retriever and reranker shown in parens (zero-indexed). Related words are highlighted.
    }
    \label{fig:error_analysis_1}
\end{figure}

\begin{figure}[th!]







\begin{subfigure}[b]{\linewidth}
\begin{tcolorbox}[colframe=black, colback=white, boxrule=0.5pt, width=\linewidth, boxsep=5pt, left=5pt, right=5pt,
title={\small\centering text-embedding-3-large + cohere-rerank-english-v3.0}]
\input{examples/cohere-gabonese}
\end{tcolorbox}
\end{subfigure}
\vspace{0.5mm}
    \caption{An unexpected reranker errors from Scifact. The ranks assigned by the retriever and reranker shown in parens (zero-indexed). Related words are highlighted.
    }
    \label{fig:error_analysis_2}
\end{figure}


\subsection{Listwise-Reranking with Large Language Models \label{sec:listwise}}

Our results in the previous subsections show that scaling $K$ can eventually hurt reranker recall and eventually yield worse performance than retrieval. It remains unclear whether this is an issue with reranking in general, or whether certain rerankers are more robust to the failure mode of scaling K. We additionally explore listwise reranking using LLMs in order to see if there is at least one class of model where reranking is more robust. Compared to cross-encoders, this new approach is advantaged primarily for two reasons: the model uses more context when reranking and leverages a strong LLM backbone, \textit{gpt-4o-mini}.

\paragraph{Implementation} We follow the sliding window approach introduced in \citet{sun-etal-2023-chatgpt} using a window size of 20, stride size of 10, and \textit{gpt-4o-mini} as the LLM.\footnote{Anecdotally, we found Llama-3.1-70B yielded similar results as gpt-4o-mini, although a full exploration of LLM performance is outside the scope of this study.}$^,$\footnote{We use sliding window with zero-shot prompting. Further performance improvements, potentially at lower inference cost, can be can be achieved through finetuning~\citet{pradeep2023rankzephyr} as well as alternative prompting algorithms~\cite{huang2024instupr,Parry2024TopDownPF}.}
When our program fails to parse the output of the LLM due to syntax or logical errors (e.g. a repeated document ID), then we simply keep the retrieval ordering for that window.\footnote{Errors like this occur about 5\% of the time when $K=100$, aligning with previous reports \cite{pradeep2023rankzephyr}.}
When inspecting the outputs from the model on failed windows, we observe that almost all of the errors occurred when the model thought that none of the documents being ranked were relevant to the query, and thus either didn't follow the format specified in the prompt or refused to output a ranking altogether.

\paragraph{Experiment}
Running the pointwise cross-encoder on the full data is relatively inexpensive compared with the listwise LLM approach. The cross-encoder scores can be reused to compute Recall@10 for all values of $K$. In contrast, since the listwise approach begins with the lowest scoring documents from retrieval, this means that there is no opportunity to reuse computation between different values of $K$. This is exacerbated by the fact that \textit{gpt-4o-mini} has a substantially higher cost per token than other rerankers.\footnote{Cost per 1M tokens (\texttt{Feb-18-2024}): \textit{voyage-2} is $\$0.05$ and \textit{gpt-4o-mini} is $\$0.15$.} For these reasons, we only run the listwise approach with one of the retrievers and only across four settings of $K \in \{50, 100, 500, 1000\}$.

\paragraph{Results}
We report results for listwise reranking with LLMs in Figures~\ref{fig:recall_10_plots} and \ref{fig:recall_10_plots_enterprise} (additional results shown in the Appendix Figures~\ref{fig:individual_recall_10_academic}, \ref{fig:individual_recall_10_academic_cont}, and \ref{fig:individual_recall_10_enterprise}). Not only is Recall@10 substantially higher than competing rerankers, but consistently improves as we scale $K$.
In contrast, we observed that pointwise cross-encoders were robust in only 23.3\% and 22.2\% of academic and enterprise experiments, respectively (see ``Helps \& Never Hurts'' in Table~\ref{tab:analysis_scaling}).
This is an encouraging proof of existence that there are available rerankers with favorable properties in respect to scaling $K$.




\section{Discussion}\label{main:discussion}


Our research has focused on empirically analyzing how effectively we can improve reranking by scaling up the number of included documents.
We primarily focus on off-the-shelf pointwise cross-encoder rerankers and show that in many cases scaling $K$ hurt performance.\footnote{
The one reranking alternative that we explore is listwise reranking with LLMs (\S\ref{sec:listwise}). The listwise approach is encouraging and serves as a proof of existence that there are reranking architectures where scaling $K$ does help. Even so, this listwise approach is high latency and impractical, and is not our focus.}
In this section, in order to aid future research, we highlight three potential causes for unexpected failures in rerankers and how they might be addressed.

\paragraph{Negatives in Training} 
We conjecture that today’s point-wise rerankers lack of robustness is partially due to \textit{exposure bias} during reranking training.
Namely, pointwise rerankers may see less negatives in training than their embedding model counterparts because large batch reranker training is computationally expensive and does not easily share negatives across the batch. Thus, we hypothesize that rerankers are trained on negatives selected from a subset of the corpus filtered by retrievers, which may explain the peak performance of rerankers occuring at a lower number of retrieved documents, where the documents the reranker scores more closely resemble its training data. Thus, while it is theoretically intuitive for rerankers to be considered strictly better models than embedding models, the computational expense of fully training a reranker is prohibitive enough to prevent this from being the case today. 

For example, when training the mGTE embedding and reranking model in \cite{zhang2024mgte}, the authors use 16,384 randomly selected in-batch negatives to train their embedding model, while only using 4 randomly selected negatives to train their cross-attention based reranker model. Additionally, some approaches to training embedding models are able to scale to millions of negative documents \cite{xiong2020approximate, lindgren2021efficient}. Under-trained rerankers may explain why previous research has shown embeddings can match cross-encoder quality \cite{menon22defense}.

\paragraph{Reranking as Ensembling} It's confusing that rerankers can help when we've shown that they are overall worse than retrievers at tasks like full retrieval (\S\ref{sec:full_rerank}). It's similarly counter-intuitive that they can be worse than retrievers given that a cross-encoder and dense embedding are essentially the same model architecture, but trained with different input data. Perhaps these observations make more sense if we view reranking as a type of ensembling.
Diversity is a useful property in ensembles \cite{Maclin1999PopularEM},
meanwhile,
rerankers are overly specialized compared with embeddings due to the aforementioned \textit{exposure bias}.
From this view, it's plausible that
the reranker is providing alternative features (compared to embeddings) that are helpful to discriminate relevant from irrelevant documents only from the top of the list.
The ensemble perspective is even more obvious when taking BM25 into account, which is case where the first stage and second stage are modeled completely differently, yet reranking remains helpful.


The ensembling perspective could be fruitful in future research. Scaling laws for dense retrieval explain how retrieval quality may improve with an increase in model parameters and training data \cite{Fang2024ScalingLF}, although they do not factor in reranking. 
\citet{Grefenstette2018StrengthIN} shows that ensembling may be more productive than training a single large model under a fixed compute budget, and the same may apply to reranking.

\paragraph{Robustness of Deep Learning}

There have been many studies that suggest deep learning models aren't robust. Representative instances of this problem include vision models failing to classify images due to small perturbations 
\cite{deeplearningvisionrobustness}, as well as text processing models changing their prediction when substituting words in their input with synonyms \cite{nlprobustness}. 
In the context of reranking, as one scales the number of documents, each additional document included creates a risk that the model may assign an inappropriately high score to an irrelevant document \cite{zamani2022stochasticrerank}.
Similar observations have been made in other AI paradigms. For example, Best-of-N decoding can suffer with large N due to \textit{reward hacking}, which leads to finding a more favorable proxy reward that doesn't necessarily improve results on the downstream task \cite{learn2summarize,Nakano2021WebGPTBQ,Pan2022TheEO,gao2023rm,Lambert2023TheAC,rafailov2024scaling,Stroebl2024InferenceSF}. 

\paragraph{Pointwise vs. Listwise Ranking} The lack of robustness in rerankers is exacerbated when they assign scores independently to each document (pointwise). Using models that are trained with a listwise loss or listwise inference procedure has been shown to improve ranking quality \cite{ai2018listwise,ai2019listwise,rahimi2019listwise,Gao2023PolicyGradientTO}. This is one explanation why listwise-reranking with large language models outperformed pointwise-reranking with cross-encoders in \S\ref{sec:listwise}.

\paragraph{Effective Finetuning}  The research community has highlighted challenges associated with finetuning. One explanation for why our listwise-reranker may outperform the cross-encoder rerankers is that the listwise-reranker is based on gpt-4o-mini without any finetuning. The rerankers, on the otherhand, have almost definitely been finetuned for the reranking task, which may have lead to catastrophic forgetting or other modeling issues. Although the shortcomings of current finetuning recipes are well established \cite{loradatabricks}, finetuning for ranking presents even further challenges since this procedure is used to convert models from text generation to embedding or classification \cite{Ma2023FineTuningLF,llm2vec}. 

\paragraph{Favorable Conditions for Rerankers} In our study, we have specifically chosen strong first stage retrieval methods (i.e. dense embeddings), as well as a diverse collection of reasonable datasets that are representative of real world workloads. Prior studies are often more limited to settings that are particularly well suited for reranking: using BM25 as the first stage retrieval and evaluated against MSMarco or similar datasets \cite{hof2019lowfreq,hof2020constrained,Lin2020PretrainedTF,mokrii2021pseudolabel,Parry2024GenerativeRF}.

We include at least one result where  reranking is consistently helpful---the BM25 portion of Figure~\ref{fig:recall_10_plots_enterprise} shows clear benefits from reranking on the Enterprise data split. Even so, we also find that \textit{dense} embeddings alone are on par or better than \textit{sparse} embeddings in the same data setting.
The different trends observed when reranking atop sparse or dense embeddings in the first stage highlights the importance of exploring many scenarios when conducting reranker research.
In practice, there are more free variables than simply the combination of dataset, retriever, and reranker. For instance, \citet{bahri2020choppy} demonstrate the benefits of adaptive ranked list truncation---although this filtering would typically be applied prior to reranking, it may be beneficial to apply afterwards as well.
\section{Conclusion}

We empirically study how, in modern IR systems, scaling the inference compute of different rerankers impacts the quality of the retrieved output. We do this by testing modern OSS and closed source embedding models and rerankers on  carefully curated academic and enterprise datasets. We find that, for modern cross-encoders, scaling the inference compute by reranking more documents ultimately leads to significant performance degredation on Recall.
Furthermore, when ranking the entire corpus, we find that modern embedding models outperform cross-encoders.
As a path forward, we present evidence that using LLMs for listwise reranking can outperform cross-encoders. This approach could serve as a potential teacher to improve cross-encoders or serve directly as a reranker when the cost-quality tradeoff is desirable.
We hope our findings and analyses will be useful resources to practitioners as they deploy reranker pipelines, as well as spur future research to improve cross-encoder rerankers.

\section*{Limitations}
In our experiments, several of the models used are closed source, where we do not have access to the information like the training data, precise model architecture, and model size. As mentioned in \S\ref{main:discussion}, experimenting with different training strategies, training data distributions, and model sizes may confirm some of our hypotheses, as well as lead to new insights regarding rerankers. 


\bibliography{references}

\clearpage

\appendix

\section{Details for Reproducibility \label{app:reproducibility}}

\subsection{Model Details}

Table~\ref{tab:model_links} includes reference links for each embedding and reranker model used in this study.

\subsection{Computational Resources}

Experiments were run on commodity GPUs, and no individual run required more than 24 hours.

\subsection{Data Preprocessing}

Summary of downsampling for datasets are in Table~\ref{table:datasets}. For the datasets, we either downsampled the corpus size, number of queries, or made gold labels stricter, which we shall explain in more detail.

\paragraph{Corpus Downsampling} When downsampling the size of the corpus, we made sure the maximum size of the corpus was $10,000$ documents. To construct this new corpus, we first added all of the gold documents from the queries into the corpus to make sure those queries could be correctly answered. Then, we randomly sampled from the larger corpus without replacement to choose the remaining documents. This corpus downsampling was done for BRIGHT's biology split. 

\paragraph{Query Downsampling} When downsampling the queries, we simply randomly selected the new queries from the main dataset. This downsampling was done for Scifact. 

\paragraph{BRIGHT} For the BRIGHT datasets, we use the `documents' split and the gemini-generated reasoning queries, as this type of query reformulation was deemed to work more effectively in the BRIGHT paper.

\paragraph{BIRCO} For the BIRCO datasets, the gold documents are densely labeled with `qrel' scores, which is used to gauge how relevant a given gold document is to the query. The higher the qrel score, the more relevant that gold document is. In order to accomodate the binarization of relevance judgement scores that recall requires, we create a new dataset such that all the new gold documents are the gold documents of the original dataset that share the highest qrel score for a given query.

\subsection{Evaluation Metrics}

We use Recall as our primary metric for evaluation because retrieval augmented generation (RAG) is increasingly becoming the main way that search engines are accessed \cite{salemi2024llmsearch}. In lieu of order-sensitive metrics such as nDCG, we include Recall@K with various values of K.

\section{Detailed Results \label{app:detailed_results}}

\begin{figure*}
    \centering
    \includegraphics[width=1\linewidth]{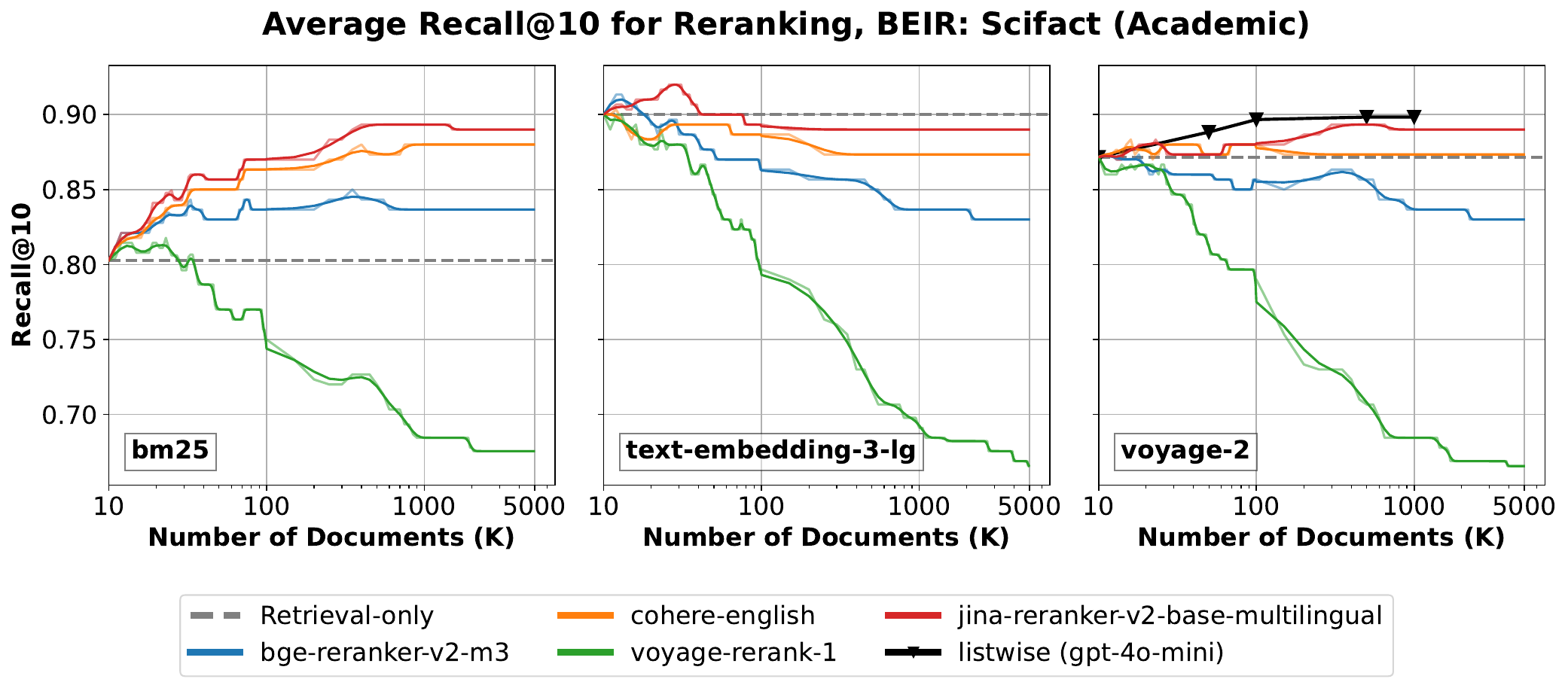} \\
    \vspace{1.5em}
    \includegraphics[width=1\linewidth]{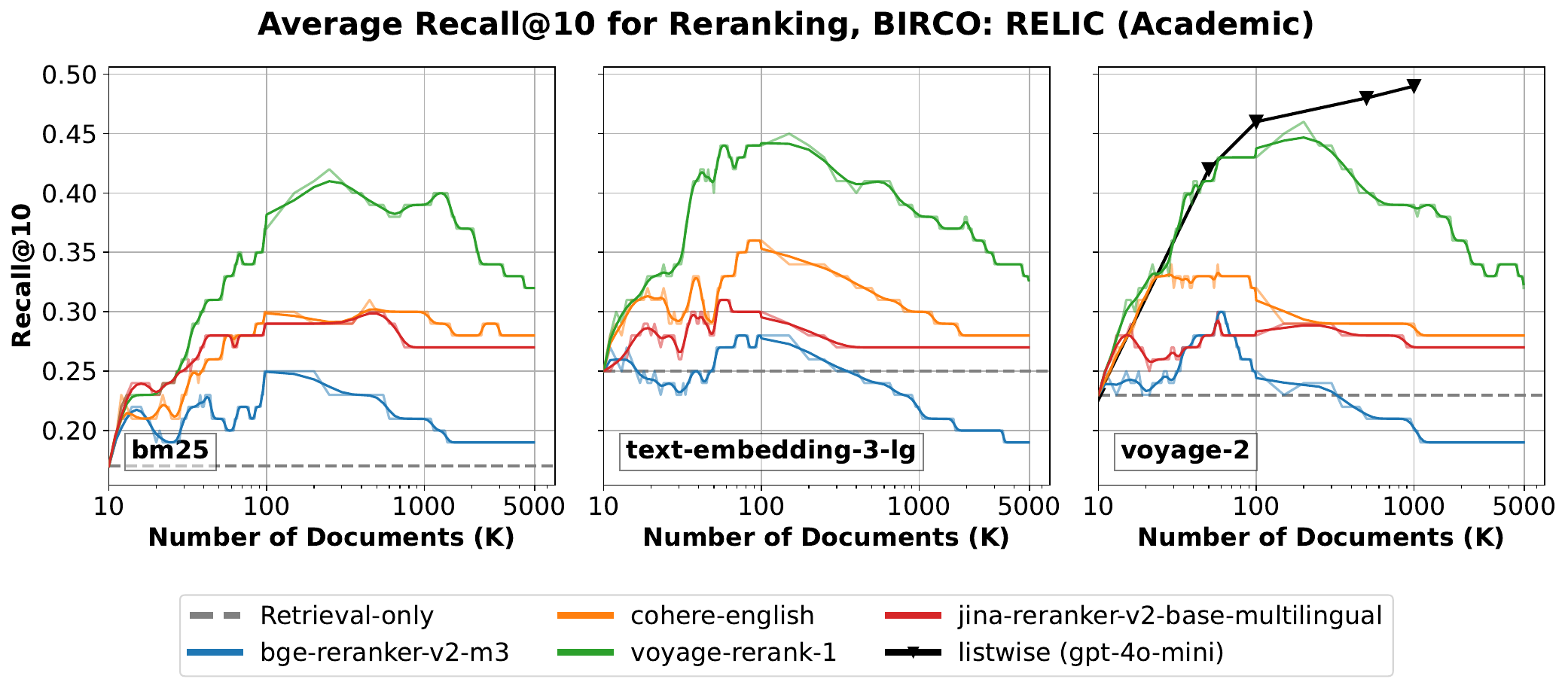} \\
    \vspace{1.5em}
    \includegraphics[width=1\linewidth]{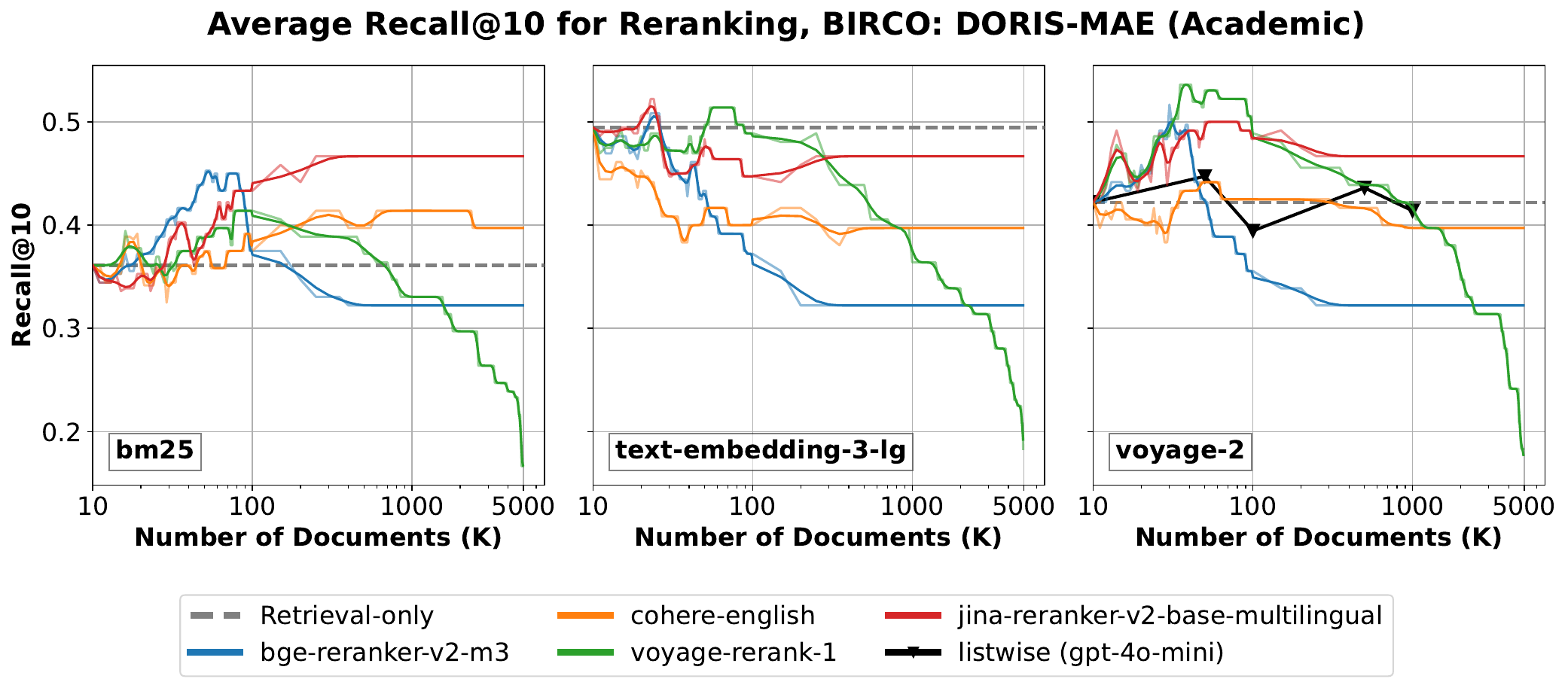}
    \caption{Recall@10 for reranking (Academic).}
    \label{fig:individual_recall_10_academic}
\end{figure*}

\begin{figure*}
    \centering
    \includegraphics[width=1\linewidth]{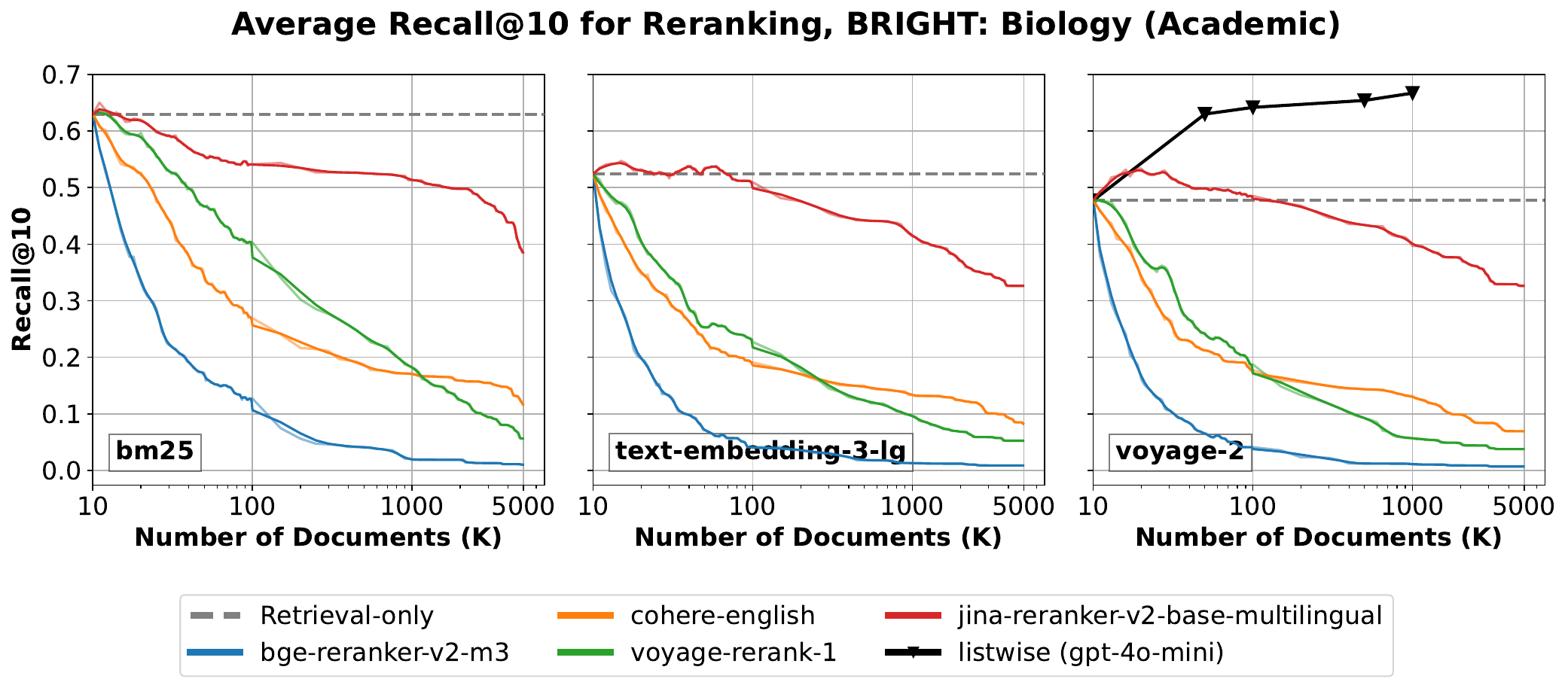} \\
    \vspace{1.5em}
    \includegraphics[width=1\linewidth]{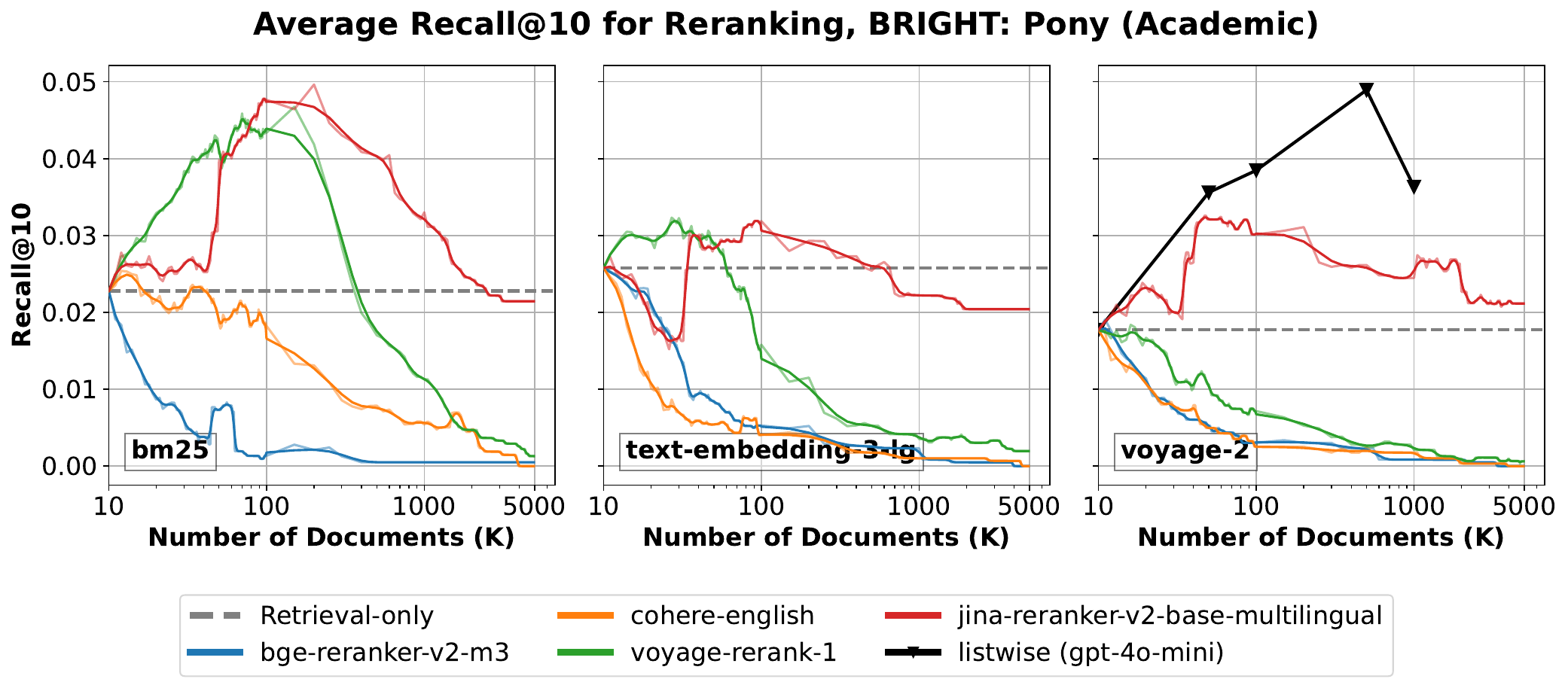}
    \caption{Recall@10 for reranking (Academic cont.).}
    \label{fig:individual_recall_10_academic_cont}
\end{figure*}

\begin{figure*}
    \centering
    \includegraphics[width=1\linewidth]{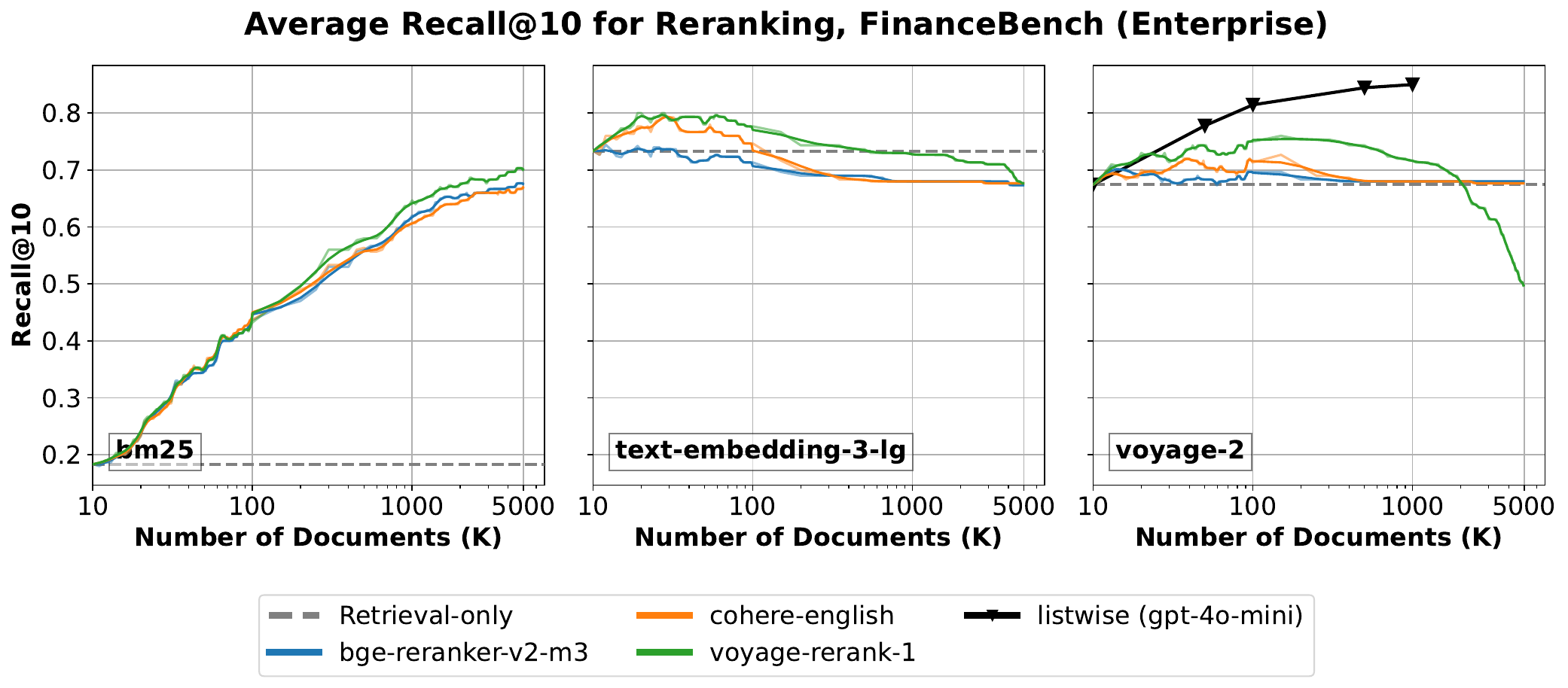} \\
    \vspace{1.5em}
    \includegraphics[width=1\linewidth]{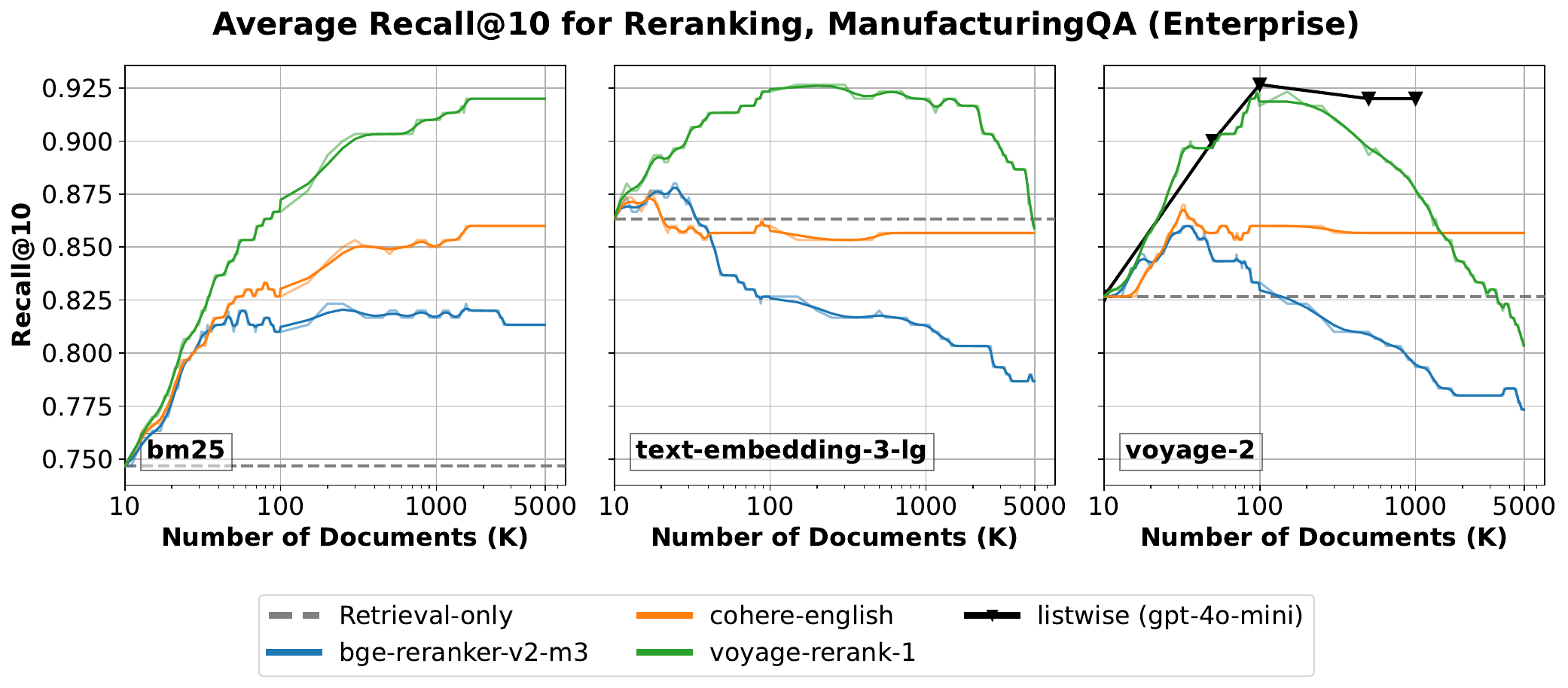} \\
    \vspace{1.5em}
    \includegraphics[width=1\linewidth]{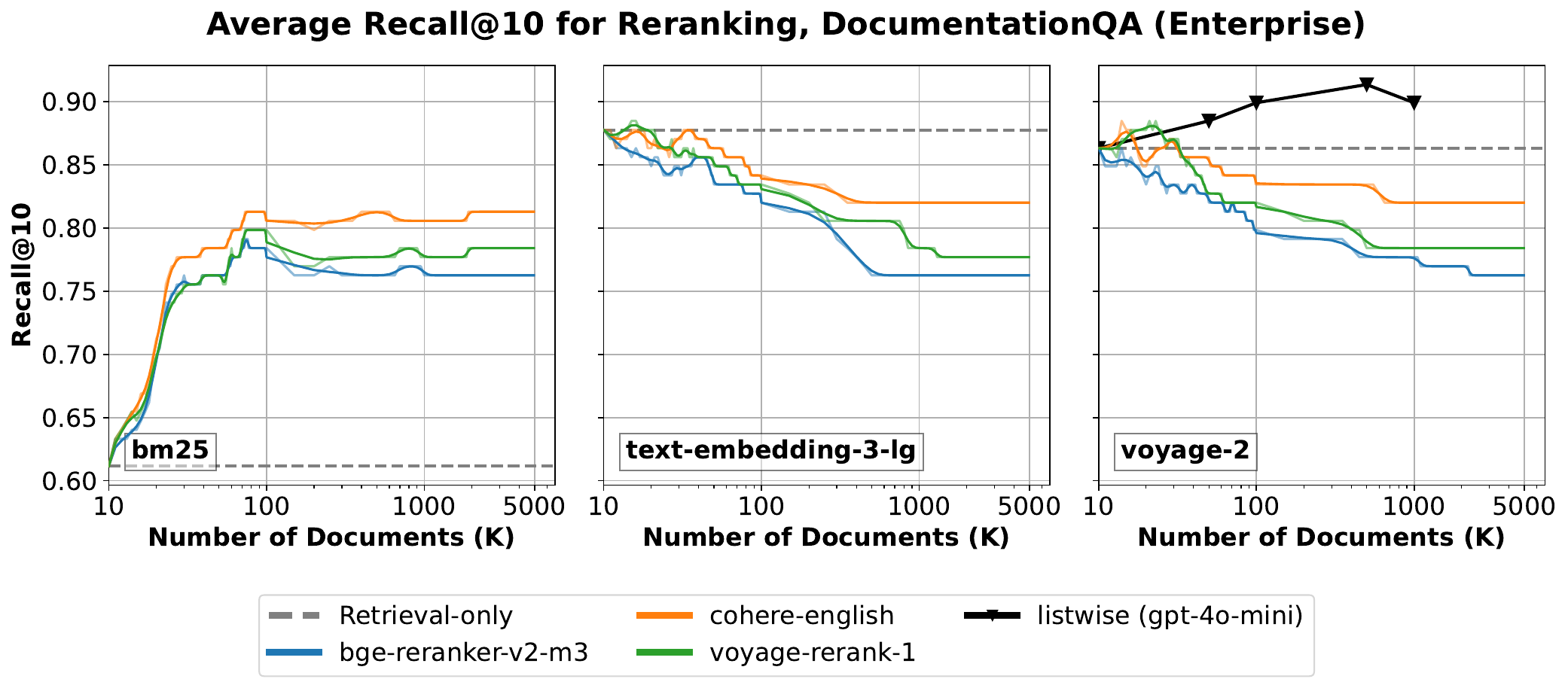}
    \caption{Recall@10 for reranking (Enterprise).}
    \label{fig:individual_recall_10_enterprise}
\end{figure*}

We show Recall@10 for individual datasets across academic and enterprise splits in Figures \ref{fig:individual_recall_10_academic}, \ref{fig:individual_recall_10_academic_cont}, and \ref{fig:individual_recall_10_enterprise}.

\section{Example Rankings \label{app:error_analysis}}

\paragraph{Reranker Failures} We present and discuss unexpected reranker errors in \S\ref{sec:error_analysis}. For the associated queries we report the top-8 retriever and reranker results in 
Figures
\ref{fig:full_query_voyage_retriever},
\ref{fig:full_query_voyage_reranker},
\ref{fig:full_query_bge_retriever},
\ref{fig:full_query_bge_reranker},
\ref{fig:full_query_jina_retriever},
\ref{fig:full_query_jina_reranker}.


\FloatBarrier

\begin{table*}[t!]
\resizebox{\linewidth}{!}{%
\begin{tabular}{|l|l|l|}
\hline
\textbf{Model Name}                         & \textbf{Type} & \textbf{Link}                                                                                                   \\ \hline
\textit{voyage-2}                           & E             & \href{https://blog.voyageai.com/2024/05/05/voyage-large-2-instruct-instruction-tuned-and-rank-1-on-mteb/}{https://blog.voyageai.com/2024/05/05/voyage-large-2-instruct-instruction-tuned-and-rank-1-on-mteb/} \\ \hline
\textit{text-embedding-3-large}             & E             & \href{https://openai.com/index/new-embedding-models-and-api-updates/}{https://openai.com/index/new-embedding-models-and-api-updates/}                                                  \\ \hline
\textit{voyage-rerank-1}                    & R             & \href{https://blog.voyageai.com/2024/05/29/voyage-rerank-1-cutting-edge-general-purpose-and-multilingual-reranker/}{https://blog.voyageai.com/2024/05/29/voyage-rerank-1-cutting-edge-general-purpose-and-multilingual-reranker/}    \\ \hline
\textit{cohere-rerank-v3}                   & R             & \href{https://cohere.com/blog/rerank-3}{https://cohere.com/blog/rerank-3}                                                                                \\ \hline
\textit{bge-reranker-v2-m3}                 & R             & \href{https://huggingface.co/BAAI/bge-reranker-v2-m3}{https://huggingface.co/BAAI/bge-reranker-v2-m3          }                                                        \\ \hline
\textit{jina-reranker-v2-base-multilingual} & R             & \href{https://jina.ai/news/jina-reranker-v2-for-agentic-rag-ultra-fast-multilingual-function-calling-and-code-search/}{https://jina.ai/news/jina-reranker-v2-for-agentic-rag-ultra-fast-multilingual-function-calling-and-code-search/} \\ \hline
\end{tabular}
}
\caption{Models, their type (E=Embedding or R=Reranker), and reference links.}
\label{tab:model_links}
\end{table*}


\begin{table*}[t!]
\centering
\begin{tabular}{|l|l|l|l|}
\hline
\textbf{Split} & \textbf{Benchmark} & \textbf{CDS} & \textbf{QDS} \\ \hline
Scifact \cite{wadden-etal-2020-fact} & BEIR \cite{thakur2021beir} & No & Yes \\ \hline
RELIC \cite{relic} & BIRCO \cite{Wang2024BIRCOAB} & No & No \\ \hline
DORIS MAE \cite{doris-mae-og} & BIRCO \cite{Wang2024BIRCOAB} & No & No \\ \hline
Biology & BRIGHT \cite{BRIGHT}  & Yes & No \\ \hline
Pony & BRIGHT \cite{BRIGHT} & No & No \\ \hline
DocumentationQA& Enterprise & No & No \\ \hline
FinanceBench \cite{financebench} & Enterprise & Yes & N/A \\ \hline
ManufacturingQA & Enterprise & No & No \\ \hline
\end{tabular}
\caption{Summary table for datasets used in our evaluation.
CDS = Corpus Downsampled. QDS = Query Downsampled.
Query details for Financebench are proprietary.
}
\label{table:datasets}
\end{table*}

\FloatBarrier

\clearpage


\begin{figure*}[th!]
\begin{tcolorbox}[colframe=black, colback=white, boxrule=0.5pt, width=\linewidth, boxsep=5pt, left=5pt, right=5pt,
title={\small\centering text-embedding-3-large + voyage-rerank-1 (retriever top-8)}]
\input{query_results/retriever-beir_scifact_150-756-text-embedding-3-large-voyage-rerank-1.tex}
\end{tcolorbox}
\caption{Top-8 results from the retriever on Scifact. The ranks assigned by the retriever and reranker shown in parens (zero-indexed). }
\label{fig:full_query_voyage_retriever}
\end{figure*}

\begin{figure*}[th!]
\begin{tcolorbox}[colframe=black, colback=white, boxrule=0.5pt, width=\linewidth, boxsep=5pt, left=5pt, right=5pt,
title={\small\centering text-embedding-3-large + voyage-rerank-1 (reranker top-8)}]
\input{query_results/reranker-beir_scifact_150-756-text-embedding-3-large-voyage-rerank-1.tex}
\end{tcolorbox}
\caption{Top-8 results from the reranker on Scifact. The ranks assigned by the retriever and reranker shown in parens (zero-indexed). }
\label{fig:full_query_voyage_reranker}
\end{figure*}

\begin{figure*}[th!]
\begin{tcolorbox}[colframe=black, colback=white, boxrule=0.5pt, width=\linewidth, boxsep=5pt, left=5pt, right=5pt,
title={\small\centering text-embedding-3-large + bge-reranker-v2-m3 (retriever top-8)}]
\input{query_results/retriever-beir_scifact_150-1200-text-embedding-3-large-bge-reranker-v2-m3}
\end{tcolorbox}
\caption{Top-8 results from the retriever on Scifact. The ranks assigned by the retriever and reranker shown in parens (zero-indexed). }
\label{fig:full_query_bge_retriever}
\end{figure*}

\begin{figure*}[th!]
\begin{tcolorbox}[colframe=black, colback=white, boxrule=0.5pt, width=\linewidth, boxsep=5pt, left=5pt, right=5pt,
title={\small\centering text-embedding-3-large + bge-reranker-v2-m3 (reranker top-8)}]
\input{query_results/reranker-beir_scifact_150-1200-text-embedding-3-large-bge-reranker-v2-m3.tex}
\end{tcolorbox}
\caption{Top-8 results from the reranker on Scifact. The ranks assigned by the retriever and reranker shown in parens (zero-indexed). A few documents are truncated to conserve space.}
\label{fig:full_query_bge_reranker}
\end{figure*}

\begin{figure*}[th!]
\begin{tcolorbox}[colframe=black, colback=white, boxrule=0.5pt, width=\linewidth, boxsep=5pt, left=5pt, right=5pt,
title={\small\centering text-embedding-3-large + cohere-rerank-english-v3.0 (retriever top-8)}]
\input{query_results/retriever-beir_scifact_150-690-text-embedding-3-large-cohere-english}
\end{tcolorbox}
\caption{Top-8 results from the retriever on Scifact. The ranks assigned by the retriever and reranker shown in parens (zero-indexed). A few documents are truncated to conserve space.}
\label{fig:full_query_jina_retriever}
\end{figure*}

\begin{figure*}[th!]
\begin{tcolorbox}[colframe=black, colback=white, boxrule=0.5pt, width=\linewidth, boxsep=5pt, left=5pt, right=5pt,
title={\small\centering text-embedding-3-large + cohere-rerank-english-v3.0 (reranker top-8)}]
\input{query_results/reranker-beir_scifact_150-690-text-embedding-3-large-cohere-english}
\end{tcolorbox}
\caption{Top-8 results from the reranker on Scifact. The ranks assigned by the retriever and reranker shown in parens (zero-indexed). }
\label{fig:full_query_jina_reranker}
\end{figure*}

\end{document}